\documentclass{epl}

\title{Two dynamic exponents in the resistive transition of fully frustrated
Josephson-junction arrays}
\shorttitle{Dynamic exponents in frustrated Josephson arrays}
\author{E. Granato\inst{1} \and D. Dom\'{\i}nguez\inst{2}}
\institute{\inst{1} Laborat\'orio Associado de Sensores e Materiais,
Instituto Nacional de Pesquisas Espaciais -
12201-190 S\~{a}o Jos\'e dos Campos, S\~ao Paulo, Brazil \\
\inst{2} Centro At\^omico Bariloche
- 8400 San Carlos de Bariloche, Rio Negro, Argentina}


\begin{document}

\maketitle

\begin{abstract}
We study the  resistive transition in Josephson-junction arrays at
$f=1/2$ flux quantum per plaquette by  dynamical simulations of
the resistively-shunted-junction model. The current-voltage
scaling and critical dynamics of the phases are found to be well
described by the same critical temperature and static exponents as
for the chiral (vortex-lattice) transition. Although this behavior
is consistent with a single transition scenario, where phase and
chiral variables order simultaneously, two different dynamic
exponents result for phase coherence and chiral order.

\end{abstract}

Phase transitions in two-dimensional Josephson junction arrays
(JJA) has been a subject of much investigation
\cite{conf,zant,carini,martinoli,yu,ling,teitel,gkn,lkg,rj,gkn96,leelee,olsson,%
korshu,wallin,tiesinga,jensen,zheng,teitel89,simkin98,leeteitel,md01,dd99,gd97,%
eg98,theron,beck,shenoy,minhagen}. Such arrays can be artificially
fabricated
\cite{conf,zant,carini,martinoli} and are also closely related to
superconducting wire networks \cite{yu,ling}. Experimentally, the
resistive transition has been the one most extensively studied
\cite{zant,carini,martinoli,yu,ling}, while theoretically several
studies of  XY models
\cite{teitel,gkn,lkg,rj,gkn96,leelee,olsson,
korshu,wallin,tiesinga,jensen,zheng,teitel89,
simkin98,leeteitel,md01,dd99,gd97,eg98}, which describe the JJA,
have been done. A significant understanding of these systems has
already been  achieved by comparing the experiments with the
theoretically predicted equilibrium critical behavior with and
without an applied magnetic field.  However, to a large extent,
dynamical critical behavior remains much less explored,
particularly in the frustrated case (finite magnetic field). It is
well known that while static critical phenomena depend on the
spatial dimensionality as well as on the symmetry of the order
parameter, the dynamic universality class
will depend upon additional properties which do not affect the
statics as, for example, conservation laws for the order parameter
\cite{hh}. Thus, testing the universality hypothesis of dynamical
critical behavior requires the study of specific dynamical models.
In JJA, the physically relevant dynamical model for the phase
dynamics has not been unambiguously identified
\cite{theron,beck,tiesinga,jensen}. One would expect that, at
least for an array of ideal tunnel junctions, the
Resistively-Shunted-Junction (RSJ) model of current flow between
superconducting grains would be a more physical representation of
the system \cite{shenoy}. In experiments, the resistive transition
in JJA is usually identified from the behavior of the
current-voltage (I-V) characteristics near the critical
temperature. The divergent correlation length determines both the
linear and nonlinear resistivity sufficiently close to the
transition. To interpret the data and determine the underlying
equilibrium transition, the scaling theory of Fisher {\it et al.}
\cite{ffh} has been widely used. For JJA at zero magnetic field,
the resistive transition is in the Kosterlitz-Thouless (KT)
universality class \cite{conf,teitel,shenoy,minhagen}. Studies of
the critical dynamics, either with Monte Carlo (MC) dynamics
\cite{wallin} or with RSJ dynamics \cite{jensen}, find a behavior
consistent with the dynamical theory of the KT transition. The
exponent of the current-voltage (I-V) relation, $V\sim I^a$, at
the transition, assuming the universal value $a=z+1=3$,
corresponds to a dynamic exponent $z=2$ in  the resistivity
scaling theory \cite{ffh}.

However, in frustrated  Josephson-junction arrays (FJJA),
corresponding to $f=1/2$ flux quantum per plaquette, besides the
phase variables, the vortex-lattice induced by the external field
introduces and additional Ising-like order parameter, the
chirality \cite{teitel}, which measures the direction of local
current circulation in  the array. The ground state consists of a
pinned commensurate vortex-lattice corresponding to an
antiferromagnetic arrangement of chiralities and vortex-lattice
melting corresponds to the chiral order-disorder transition. As a
consequence, two distinct scenarios for the phase transition have
been proposed \cite{gkn,lkg,rj,gkn96,leelee,olsson,korshu}: separated
phase-coherence and chiral transitions  or a single transition
where phase and chirality order simultaneously in a different
universality class. Since the resistive transition corresponds to
the onset of phase coherence, the later scenario where the
critical dynamics should involve coupled variables may lead to
important consequences for the resistivity scaling near the
transition that can be detected experimentally. Previous numerical
studies of the I-V characteristics, obtained either with RSJ
dynamics \cite{teitel89,simkin98} or with MC dynamics
\cite{leeteitel}, were performed only for small system sizes
($L\le 16$); while other works have studied the short-time
dynamics of chirality \cite{zheng}, and the non-equilibrium
transitions at large currents \cite{md01}. In particular, the
studies with RSJ dynamics used free boundary conditions to impose
a driving current. This leads to significant additional
dissipation due to boundary effects \cite{gks98}, specially in
small system sizes. In this letter, we study in detail the
critical dynamics and resistivity scaling in FJJA by numerical
simulation of the RSJ dynamics with periodic boundary conditions
\cite{dd99,gd97} including large systems sizes ($L=64,128$). This
allowed us to find the following remarkable and unexpected results
for the dynamical properties: (i) the current-voltage behavior is
well described by a resistive transition corresponding to the
chiral transition; (ii) two different dynamic exponents,
$z_{XY}\sim 1$ and $z_{ch}\sim 2$, are found for phase and chiral
variables, respectively; and (iii) at the transition, the exponent
of the I-V power-law, $V\sim I^a$, is $a = z_{XY}+1 \approx 2$
rather than $a=3$ as for the unfrustrated case. We discuss an
interpretation of these results within the single transition
scenario including  a multicritical point and also suggest possible
explanations within the double transition picture.

We consider  a square two-dimensional array described by the
overdamped RSJ model with current conservation at each node
\cite{shenoy}. The equations of motion for the phases $\theta
_{i}$ of the superconducting order parameter located at node $i$
of the square lattice can be written as
\begin{equation}
\frac{\hbar}{2eR_{o}}\sum_{j}(\dot{\theta}_{i}-\dot{\theta}%
_{j})=-\sum_{j}[I_{o}\sin (\theta _{i}-\theta _{j}-A_{ij})+\eta
_{ij}]
\end{equation}
where $R_{o}$ is a uniform shunt resistance, $\eta _{ij}(t)$ is a
thermal noise with correlations
$\langle\eta_{ij}(t)\eta_{kl}(t^{\prime})\rangle =2k_{B}T/R_{o}
\delta_{ij,kl}\delta(t-t^{\prime})$, $I_{o}$ is the junction
critical current, $A_{ij}=\int_i^j{\bf A}.d{\bf l}$ is the line
integral of the vector potential, and $\sum_{ij}A_{ij}=2\pi f$,
with the sum taken around each elementary plaquette of the
lattice. Dimensionless quantities are used with time in units of
$\tau =\hbar
/2eR_{o}J_{o}$, current in units of $I_{o}$, voltages in units of $%
R_{o}I_{o} $ and temperature in units of $\hbar I_{o}/2ek_{B}$. A total
current $I$ is imposed uniformly in the array using fluctuating periodic
boundary conditions \cite{dd99,gd97} with current density $J=I/L$,
where $L$ is the system
size and the average electric field $E$ is obtained from the voltage $V$
across the system as $E=V/L=(\hbar/2e)\langle{d\Theta/dt}\rangle$,
with $\Theta$ the global phase difference or twist \cite{dd99}.
We integrate the dynamical equations with a second order
Runge-Kutta-Helfand-Greenside method with
time step $\Delta t=0.07\tau $, taking $10^6$ steps for the lowest
currents while for large currents $2\times 10^4$ steps were enough
for proper equilibration. The
results were averaged over $5-10$ different initial configurations
of the phases and  sizes  of $L=8 $ to $L=128$ were
considered.

\begin{figure}
\onefigure[bb= 1cm  17cm  20cm   27cm, width=12cm]{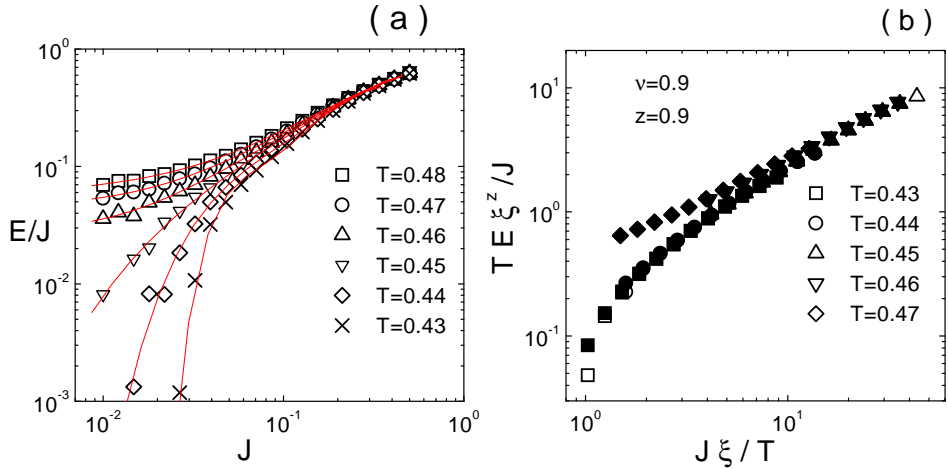}
\caption{ (a) Nonlinear resistivity $E/J$ as a function of
temperature for system size $L=64$. (b) Scaling plot of the data
for the smallest current densities. Open symbols correspond to
$L=64$ and filled ones to $L=128$. } \label{fig1}
\end{figure}

Fig. 1(a) shows the temperature dependence of the nonlinear
resistivity $E/J$ near the chiral transition temperature,
estimated previously from equilibrium Monte Carlo simulation
\cite{gkn}, $T_{ch}=0.455$. Qualitatively, the linear resistance
$R_{L}=\lim_{J\rightarrow 0}E/J$, tends to a finite value at high
temperatures but extrapolates to very low values at lower
temperatures, independent of system size, consistent with the
existence of a resistive transition  in the range $T_{c} = 0.44$
to $0.46$. In the double transition scenario, where the
phase-coherence transition is expected to be in the KT
universality class, the  estimate of the KT critical temperature
$T_{KT}=0.446$ from Monte Carlo simulations \cite{olsson}, is very
close to $T_{ch}$ and so without further analysis the true
resistive critical temperature, at $J=0$ current drive, could be
consistent with both estimates. However, we note that the IV curve
for $T_{ch}> T=0.45 > T_{KT}$ tends to zero resistivity for
$J\rightarrow 0$, while the IV curve for $T=0.46 > T_{ch}$ tends
to finite resistivity for  $J\rightarrow 0$. This already suggests
that the resistivity transition occurs at $T_{ch}$ rather than at
$T_{KT}$. In any case, the asymptotic critical behavior can be
inferred more adequately from a scaling analysis of the nonlinear
resistivity. According to the scaling theory \cite{ffh},
measurable quantities scale with the diverging correlation length
$\xi $ and the relaxation time $\tau \propto \xi ^{z}$, near the
transition temperature, where $z$ is the dynamical critical
exponent. Then, the nonlinear resistivity should satisfy the
scaling form
\begin{equation}
T\frac{E}{J}=\xi ^{-z}g_{\pm }(\frac{J}{T}\xi )  \label{scaling}
\end{equation}
in two dimensions, where the $+$ and $-$ correspond to the
behavior above and below the transition, respectively. For a
transition in the KT universality class, the correlation length
should diverge exponentially as $\xi \propto \exp (b
/|T/T_{c}-1|^{1/2})$, while otherwise  a power-law behavior is
expected $\xi \propto |T/T_{c}-1|^{-\nu }$, with an exponent $\nu
$ to be determined. Thus, a scaling plot according to Eq.
(\ref{scaling}) can be used to verify the dynamic scaling
hypotheses and the assumption of an underlying equilibrium
transition. Such scaling plot  is shown in Fig. 1(b), in the
temperature range closest to $T_{ch}$ and smallest current
densities, assuming the correlation length $\xi $ has a power-law
divergence with $T_{c}=T_{ch}$ and using $\nu$ and $z$ as
adjustable parameters so that the best data collapse is obtained.
Similar scaling analysis assuming a KT correlation length and
fixing $T_c$ at the estimates of $T_{ch}$ or $T_{KT}$ do not
result in a good data collapse \cite{gdunp}. From this scaling
analysis, we estimate $\nu=0.9(1)$ and the dynamical critical
exponent $z=0.9(2)$. We note that the static exponent $\nu$ is
consistent with estimates of the chiral transition from
equilibrium Monte Carlo simulations \cite{gkn}. However, in
simulations with MC dynamics \cite{leeteitel} a dynamic exponent
$z\sim 2$ was found, but this may correspond to a different
dynamics and also note that small systems with $L=8-14$ were
analyzed. As it is well known, finite-size effects are very
important sufficiently close to the transition when the
correlation length $\xi$ reaches the system size $L$. Previous
simulations with RSJ dynamics \cite{teitel89,simkin98} considered
very small sizes ($L=6-16$). In our case, as shown in the Fig.
1(b),  the two largest system sizes $L=64$ and $L=128$ give the
same data collapse and so finite size effects are not dominant for
this range of temperatures and current densities in our data.
Although the above scaling analysis for large system sizes already
suggests that the resistive transition temperature $T_c$ is very
close to $T_{ch}$ with $z < 2$, in absence of a completely
satisfactorily determination of $T_c$ from static critical
behavior \cite{rj,gkn96,olsson}, from now on, we will assume
\cite{commenttc}  $T_c=T_{ch}$ and explore to which extent this
give us consistent results from dynamics, including finite-size
effects. At $T_c$, the correlation length $\xi$ will be cut off by
the system size in any finite system. From Eq. (\ref{scaling}),
the nonlinear resistivity at $T_c$ should then satisfy the scaling
form
\begin{equation}
T\frac{E}{J}=L^{-z}g(\frac{J}{T}L )  \label{finitesize}
\end{equation}
We have tested the scaling form of Eq.(\ref{finitesize}) at
$T_c=T_{ch}$ for different  sizes $L=8-128$ and  we found a
very good finite size scaling with the same dynamic exponent
$z=0.9(1)$ as shown in Fig. 2(a). On the contrary, when the I-V
curves were calculated at $T=T_{KT}$  \cite{gdunp}, it was not
possible to obtain a reasonable data collapse using
Eq.(\ref{finitesize}) for $z$ values in the range  $[0.5,4]$.

\begin{figure}
\onefigure[bb= 1cm  18cm  20cm   27cm, width=12cm]{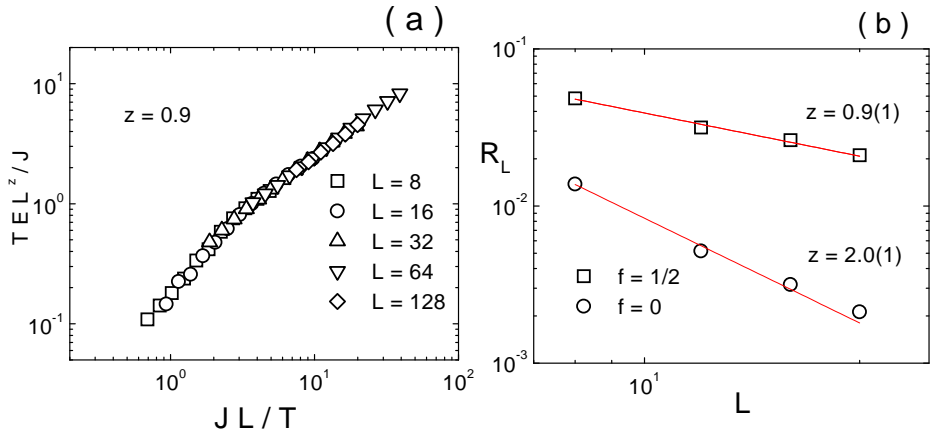}
\caption{ (a) Scaling plot of the nonlinear resistivity $E/J$ at
$T_c=T_{ch}=0.455$ for different system sizes $L$. (b) Linear
resistance as a  function of system size at the critical
temperatures $T_c=T_{ch}$ for $f=1/2$ and $T_c=0.887$ for $f=0$.
Power-law fits give estimates exponent $z$.} \label{fig2}
\end{figure}

It could still be argued that our estimate of $z$ is based on a
scaling analysis of the nonlinear I-V characteristics, which is a
nonequilibrium property, and so does not reflect the underlying
equilibrium  transition. However, additional equilibrium
calculations of the linear resistance at $T_{ch}$ give a
consistent estimate. From Eq. (\ref{finitesize}), the linear
resistance $R_{L}=\lim_{J\rightarrow 0}E/J$ at $T_c$ should  scale
as $R_L \propto L^{-z}$. The  linear resistance can be obtained
from the Kubo formula of equilibrium voltage fluctuations  as
\begin{equation}
R_L=(1/2T) \int d t \langle V(t) V(0) \rangle ,
\end{equation}
without an imposing driving current, and can also be  conveniently
determined from the long-time fluctuations of the phase difference
across the system \cite{eg98}. Fig. 2(b), shows the finite size
behavior of $R_L$. A power-law fit gives $z=0.89(6)$ which is in
fact consistent with the estimate from the I-V scaling and
suggests therefore that the value of $z$ corresponds to the
underlying equilibrium dynamical behavior. For comparison, it is
also shown in Fig. 2(b) the behavior for the unfrustrated case,
$f=0$. In this case the resistive transition is in  KT
universality class and  a dynamical exponent  $z=2$ is expected,
independent of the particular dynamics \cite{wallin,jensen}.
Indeed, for $f=0$, the same power-law fit at the critical
temperature $T_c=0.887$ estimated previously from Monte Carlo
simulations \cite{weber} gives $z=2.0(1)$.

\begin{figure}
\onefigure[bb= 1cm  8cm  19cm   19cm,width=6cm]{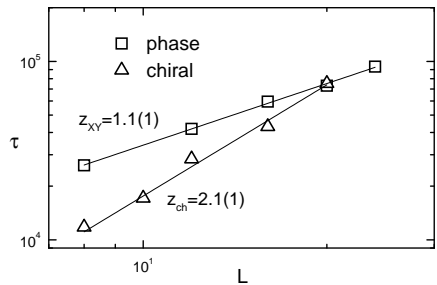}
\caption{Finite-size behavior of the phase and chiral relaxation
times,  $\tau_{XY}$ and $\tau_{ch}$ respectively,  at the critical
temperature $T_c=T_{ch}$. Power-law fits give estimates of the
dynamical exponents $z_{XY}$ and $z_{ch}$. } \label{fig3}
\end{figure}

Finally, to further verify that the estimate of $z$ does in fact
reflect critical phase fluctuations near the transition, we have
also performed equilibrium calculations of  the phase
autocorrelation function $C_{XY}(t)=\langle\vec S(0) \cdot \vec
S(t)\rangle$, where $\vec S=\sum_i \vec s_i$ and $\vec s
=(\cos(\theta),\sin(\theta))$. The relaxation time $\tau_{XY}$ can
be obtained from the exponential decay $C_{XY}(t) \propto
\exp(-t/\tau_{XY})$ at long times. Similar calculations were also
performed for the chirality autocorrelation function $C_{ch}(t)$
to obtain $\tau_{ch}$, with the local chirality defined as $\chi =
\sum_{<ij>} (\theta_i -\theta_j - A_{ij}) /2\pi$, where the
summation is taken over the elementary plaquette of the lattice
and the gauge-invariant phase difference is restricted to the
interval $[-\pi,\pi]$. From dynamic finite-size scaling, the
relaxation time should scale at $T_c$ as $\tau \propto L^z$, from
which the $z$ can be estimated from the slope in a loglog plot.
Fig. 3 shows the finite-size behavior of the relaxation time at
$T_{ch}$ for the phases and chiralities. From a power-law fit we
obtain $z_{XY}=1.1(1)$ from the phase relaxation time $\tau_{XY}$
which is indeed consistent with the estimate of $z$ from the
resistivity scaling discussed above. Naively, if the two
transitions happen at the same temperature, one would expect that
the same dynamic exponent should also hold for the chiral
relaxation time. Surprisingly, however, the estimate from the
chiral relaxation time in Fig. 3 is quite different,
$z_{ch}=2.1(1)$. Nevertheless, this value  is consistent with the
result of short-time dynamics obtained from MC simulations
\cite{zheng}.

Thus, our numerical results imply a single transition at $f=1/2$
with two dynamic exponents, $z_{XY}\sim 1$ and $z_{ch}\sim 2$. It
should be noted that two dynamic exponents at the transition does
not necessarily imply a breakdown of dynamic scaling in the
restricted sense \cite{hh}, since only the chirality is an order
parameter which develops true long-range order below the
transition. In two-dimensions, the phases can only develop
quasi-long-range order. Since $z_{ch}> z_{XY}$, the longest
relaxation time is still determined by the order parameter.
However, extended dynamic scaling \cite{hh}  still applies to the
resistivity and a different dynamic exponent is possible. Whether
this only holds for the dynamics of RSJ model studied here or also
for other dynamical models will require further work. However, we
should mention that different dynamic exponents for coupled order
parameters have already been found previously at multicritical
points in magnetic systems \cite{huber}. This would suggest that a
possible explanation for two dynamic exponents at the transition
of the FJJA may rely on the existence of a multicritical point in
the phase diagram of the relevant effective model describing the
transition. Interestingly enough, the coupled XY-Ising model
\cite{gkn} which is expected to describe the static critical
behavior of the FJJA, does have a multicritical point  and could
be a useful framework  for future investigations of the dynamical
universality class of FJJA.

Although the single transition scenario provides a consistent
interpretation of our data, we should emphasize that the
alternative separated transitions scenario can not be ruled out.
In fact, recent work by Korshunov \cite{korshu} argues for this
scenario on good theoretical grounds. If this picture turns out to
be the correct one then, we believe, there are two possible
explanations for our findings: 1) the KT transition is actually
much closer to $T_{ch}$ than estimated previously and so the
transitions can not be resolved within the accuracy of our data;
2) the scaling theory of \cite{ffh} is not valid for the present
case and should therefore be enlarged to include the interplay of
two divergent length scales at nearby temperatures
\cite{olsson,korshu} which can lead to crossover effects at small
length and time scales. However, it remains unclear how the
resistivity scaling can be so well described by the chiral
transition temperature including data for different temperature
and systems sizes.

In conclusion, we find that the resistivity scaling and critical
dynamics of FJJA are well described by the critical temperature
corresponding to the chiral (vortex-lattice) transition. Two
dynamic exponents, $z_{XY}\sim 1$ and $z_{ch}\sim 2$, are found
for phase-coherence and chiral order, respectively, and, at the
transition, the exponent of the I-V power-law, $V\sim I^a$, is $a
= z_{XY}+1 \approx 2$ rather than $a=3$ as for the unfrustrated
case.
One implication of these results for transport experiments is that
the usual method of locating the critical temperature from the
value corresponding to a nonlinear I-V exponent $a=3$, will lead
to a significant underestimate \cite{comment}. It is worth
mentioning that some experiments in overdamped arrays
\cite{carini} find a value of $a \approx 2$ at $T_c$ in agreement
with our results. In addition, resistivity scaling of experimental
data on wire networks \cite{ling} is also consistent with
power-law correlation as found here but with $z \sim 2$. Since the
dynamics of wire networks are different than the RSJ dynamics,
this value of $z$ may be the result of a different dynamical
universality class. Further detailed I-V measurements combined
with magnetic properties, which could in principle probe the
chiral transition, are needed to test our results.

\acknowledgments

The work of E.G. was supported by FAPESP and of
D.D. by CONICET and ANPCyT (PICT99 03-06343).


\begin{thebibliography}{0}
\bibitem{comment}
The experiments in {\sl underdamped} arrays of \cite{zant} were
fitted using the
KT theory and obtained $a > 3$. This leads in their case
to an estimate of the critical temperature of $T_c\approx 0.3$,
which is significantly lower than the
theoretical $T_c \approx 0.45$.
\end{thebibliography}
\end{document}